# Sentiment Analysis of Comments on Rohingya Movement with Support Vector Machine

Hemayet Ahmed Chowdhury
Department of CSE
Shahjalal University of Science and Technology
Sylhet, Bangladesh

Tanvir Alam Nibir
Department of CSE
Shahjalal University of Science and Technology
Sylhet, Bangladesh

Md. Saiful Islam
Department of CSE
Shahjalal University of Science and Technology
Sylhet, Bangladesh

*Abstract*— *The Rohingya Movement and Crisis caused a huge uproar in the political and economic state of Bangladesh. Refugee movement is a recurring event and a large amount of data in the form of opinions remains on social media such as Facebook, with very little analysis done on them.To analyse the comments based on all Rohingya related posts, we had to create and modify a classifier based on the Support Vector Machine algorithm. The code is implemented in python and uses scikit-learn library. A dataset on Rohingya analysis is not currently available so we had to use our own data set of 2500 positive and 2500 negative comments. We specifically used a support vector machine with linear kernel. A previous experiment was performed by us on the same dataset using the naïve bayes algorithm, but that did not yield impressive results.*

## I. Introduction

This analysis was created to analyse the sentiment of large amounts of text based on a specific topic. We did not want to create a system to analyse pointless data, so we tried to be as purposeful with this project as possible. The choice of topic was the Rohingya crisis and movement of recent times. We chose this specific topic because not only was it a large concept of controversy in Bangladesh, but also a recurring event that happens all over the world: refugees moving from one country to another. Therefore, we created this classifier to analyse large amounts of opinions from a nation's citizens to check what percentage of them welcome refugees inside the country (approve) and what percentage do not want them (disapprove). Obviously, the source of this opinion was social media and we chose to focus on facebook, along with a significant number of vital Rohingya related posts. Hopefully, the project can be of some future purpose for government or academic purposes, as it holds an algorithm tuned to analyse political issues and, most importantly, a labelled training corpus which can be used to train systems that make the use of neural networks or support vector machines. The current application of the project is to analyse large sums of data and classify them into the two categories of approval and disapproval.

## II. Related Work

Numerous examination works are done on Sentiment Analysis of English. Cui et al. [2] dealt with online item audits. They arranged the audits to two noteworthy classes: positive and negative. They considered around 100k item audits from diverse sites. Jagtap et al. [3] connected Support Vector Machine (SVM) and Hidden Markov Model (HMM). Their cross breed characterization model to remove the conclusion of educator criticism evaluation performed well. Alm et al. [4] Separated seven passionate words to three polar classes of positive passionate, negative enthusiastic and impartial. They utilized Winnow parameter tuning approach and got 63% precision. Agarwal et al. [5] connected unigram, tree model and highlight based model to remove twitter assumption. Unigram display is outflanked by tree model and highlight based model. The exactness they got is around 61%. Zou et al. [6] presented a model of learning bilingual word embeddings from an extensive and unlabeled dataset. They demonstrated that their model beats baselines in semantic closeness of words. Turian et al. [7] chipped away at Brown groups, embeddings of Collobert and Weston (2008) and various leveled log-bilinear embeddings. Chen et al. [8] proposed some methodologies that can separate the discharged word embeddings models. They demonstrated that embeddings can identify astounding semantics of the sentences even without having the structure. Tang et al. [9] presented a system of social affair both relevant and estimation data of the words by learning Sentiment-Specific Word Embedding. They connected their model to remove twitter feeling. The precision they got is around 83%. Demand et al. [10] Worked on skip-gram show with negative inspecting of Mikolov et al.(2013) and summed up it. They separated reliance based settings and demonstrated that they deliver diverse sorts of similitudes.

A couple of research works are done on Bengali which is noted in details in a comprehensive study on sentiment of Bengali text[11]. Chowdhury et al. [12] Applied Support Vector Machine (SVM) and Maximum Entropy (MaxEnt) to distinguish the supposition of Bengali microblog posts. They tested these two strategies by joining them with various kinds of highlights. Hasan et al.[13] portrayed a method of identifying the notions of Bengali messages by Contextual Valency Analysis. Islam et al. [14], Hossain and Dey took an approach using Naïve Bayes classification model for Bengali Language. There a supervised classification method is used with language rules for detecting sentiment for Bengali Facebook Status.

Al Amin et al. [15], Islam et al. and Uzzal et al. worked on word embedding with Hellinger PCA to detect the sentiment of Bengali comments. Word co-occurrence matrix is constructed with skip gram to determine the contextual information of the comments and sliding windows are created to gather similar words in the windows. They also took a new approach[16] of sentiment classification of Bengali comments with word2vec and Sentiment extraction of words are presented. Combining the results of word2vec word co-occurrence score with the sentiment polarity score of the words, the accuracy obtained is 75.5%. Mahmud et al. [17], Mohaimen, Islam and Jannat took an approach to predict movie success by analyzing public sentiments with a support vector machine and statistical reasoning. They used non linear RBF kernel for our sentiment classifier which achieved better accuracy than the classifiers that use linear kernels in the famous IMDB Movie Review Dataset (89.51% accuracy) and also in the Pang and Lee Movie Review Dataset (86.86% accuracy). Using our system they can predict whether a movie will be successful or not with an accuracy of 90.3%

### III. METHODOLOGY

For this analysis, we worked on two classes, Approval and Disapproval. Approval, in this context, means that the comment approves of the Rohingya people movement inside the country, or at least approves of the Rohingya community as human beings with all the human rights.Disapproval, in this context, means that the comment does NOT approve of the Rohingya community and intends to display a vibe of hatred or dislike towards their well being.Initially we tried to classify the comments with a Naive-Bayes classifier but the accuracy and precision of the classifier was less than impressive. That was probably because the context of our comments were sufficiently complex as they were very political and incredibly sentence-structure based. Another reason might have been because for the Naive-Bayes classifier, as N-grams, we used uni grams. Since the sentence structure role played a vital part in the sentiment of the comments, a naive bayes algorithm with uni grams did not really give us satisfactory results, with an accuracy of only 67%.e.g "The Rohingya muslims are terrorists" was classified as a negative sentiment for the approval of the Rohingya movement while "The Myanmar army are the terrorists" was also classified as a negative sentiment.We approached to solve this problem by completely changing our algorithm to a Support Vector Machine which uses uni-grams and bi-grams. The results were, by a big margin, more accurate.The training corpus for this project was arguably the biggest difficulty we faced as we did not find any free open source training corpus based on the Rohingya conflict on the net. Consequently, we had to create, download and label our own training corpus of 5000 comments and use it as our training corpus.

### IV. EXPERIMENTAL SETUP AND RESULTS

We have applied a set of pre-processing steps to make the comments suitable for the SVM algorithm and improve performance. The following pre-processing has been done on the comments:

i. Lower Case - Convert the comments to lower case
ii. URLs - Convert www.* or https?://* to 'URL'
iii. @username - Convert username to '__HANDLE'
iv. #hashtag - Hash tags can give us some useful information, so we replace them with the exact same word without the hash. E.g. #Apple replaced with 'Apple'
v. Trimming the comment
vi. Repeating words: People often use repeating characters while using colloquial language, such as "I'm happyyyyy". We replace characters repeating more than twice with just two characters, so that the result for above would be "I'm happyy"
vii. Emoticons: Use of emoticons is prevalent in comments. We identify a set of emoticons and replace them with the representative sentiment i.e. '**positive**' or '**negative**'. E.g. ':)' is replaced by '**positive**'. Positive refers to approval and negative refers to disapproval. Further, if emoticon(s) are found in the comments, then the SVM classifier is not called and the comment is classified as positive or negative simply based on the emoticon.Stemming algorithms are used to find the "root word" or stem of a given word. We have used the PorterStemmer.Tuning of parameters was done to improve the performance of the SVM classifier. The following parameters are found to give the best results on the cross validation set (20% of the Training Corpus) without compromising much on the speed.

i. TfidfVectorizer:

    min_df=5,
    max_df=0.95,
    sublinear_tf = True,
    use_idf = True,
    ngram_range=(1, 2)

ii. Linear SVC:

    C=0.1

The algorithm achieves an overall precision, recall and f1-score of 0.79 (79%). The details can be found in table below (can be reproduced by running training.py):

|  | precision | recall | f1-score | support |
|---|---|---|---|---|
| 0 | 0.78 | 0.79 | 0.78 | 4936 |
| 1 | 0.81 | 0.81 | 0.81 | 4936 |
| avg/total | 0.79 | 0.80 | 0.79 | 9872 |

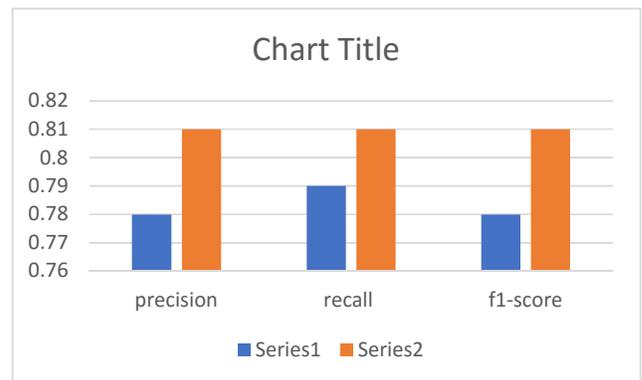

Fig: Bar Chart for Results on Two Experiments Performed

## V. CONCLUSION AND FUTURE WORK

From the experiment , we took 20% of the data set as testing data and we found that the Support Vector Machine yields an accuracy of up to 79% in analysing context based comments, which the Naïve Bayes fails to do as it cannot really take the context into account with Uni grams as features. We did not however take any specific features and even left out the word position in the sentences which caused accuracy loss as the algorithm struggled to find context. The data set was considerably small and our approach classifies the comments in only two categories. More classification categories, such as the neutral category, can be created for a better analysis. This thesis creates a platform to analyse political data and future work can be done to improve the aforementioned problems.